# Smart Drug-Delivery Systems for Cancer Nanotherapy

Paola Sánchez-Moreno[a,b], Juan Luis Ortega-Vinuesa[b], José Manuel Peula-García[b,c], Juan Antonio Marchal[d,e] and Houria Boulaiz[d,e,*]

[a]Cell Biology Unit. IRCCS San Martino University Hospital—IST National Cancer Research Institute Largo Rosanna Benzi, 10, 16132 Genova, Italy;
[b]Biocolloid and Fluid Physics Group, Department of Applied Physics, University of Granada, 18071 Granada, Spain;
[c]Department of Applied Physics II, University of Málaga, 29071 Málaga, Spain;
[d]Biopathology and Regenerative Medicine Institute (IBIMER), Centre for Biomedical Research, University of Granada, Granada E-18100, Spain;
[e]Department of Human Anatomy and Embryology, Biosanitary Institute of Granada (ibs.GRANADA), University Hospitals of Granada-University of Granada, Granada E-18012, Spain

ABSTRACT

*Background*: Despite all the advances achieved in the field of tumor-biology research, in most cases conventional therapies including chemotherapy are still the leading choices. The main disadvantage of these treatments, in addition to the low solubility of many antitumor drugs, is their lack of specificity, which leads to the occurrence of severe side effects due to nonspecific drug uptake by healthy cells.
*Objective*: The purpose of this manuscript is to review and analyze the recent progress made in cancer nanotherapy.
*Results*: Progress in nanotechnology and its application in medicine have provided new opportunities and different smart systems. Such systems can improve the intracellular delivery of the drugs due to their multifunctionality and targeting potential. First, we provide a global overview of cancer and different smart nanoparticles currently used in oncology. Then, we analyze in detail the development of drug-delivery strategies in cancer therapy, focusing mainly on the intravenously administered smart nanoparticles. Finally, we discuss the challenges, clinical trials, marketed nanomedicines and future directions of the nanotherapy applied to cancer treatment.
*Conclusion*: In this review, we have evidenced the tremendous potential that smart drug-delivery systems have to enhance the therapeutic effect of current standard treatment modalities, including chemotherapies and radiotherapies.

## 1. INTRODUCTION

Nanotechnology is an emerging science involved in the design, characterization, and manipulation of devices, structures, and systems by controlling size and shape at atomic, molecular, and supramolecular level [1]. Progress in nanotechnology has resulted in the development of novel nanomaterials with physico-chemical characteristics (higher surface-to-volume ratio) that make them an excellent candidate to be applied in the biomedical science. The applications in the screening, diagnosis, and treatment of diseases, are referred to as "nanomedicine", a new discipline of science and engineering that has the ability to dramatically change individual and population-based health care [2]. While the basic approach of conventional therapies is to eliminate diseased cells faster than healthy cells, nanomedicine attempts to use sophisticated approaches and integrates innovations in genomics and proteomics heading toward a more personalized medicine in order to improve treatment for many diseases [3, 4].

The use of nanoparticles allows enhancing our under- standing of diseases pathogenesis and providing treatment at a molecular level. Currently, nanomedicine is playing an important role in the development of the pharmaceutical in- dustry, either in the form of drugs nanoparticle-based delivery systems or imaging agents, connecting a broad range of disciplines —engineering, biology, physics, and chemistry— and leading to numerous publications and patents [see (Fig. 1)]. However, a discrepancy does exist between the number of nanomaterials in clinical use and the number of published papers in cancer nanomedicine [5]. Indeed, in the field of cancer therapy, many of the nanomedicines being investigated today have improved solubility and have tailored the release rates to enhance the pharmacokinetics of the drugs (such as platinates, camptothecins, and adriamycins) and to reduce their adverse side effects.

In fact, only 5% of the classical low-molecular-weight drugs, that reach the clinical trial stage ever make it to the market [6]. This is likely because, reformulating the currently approved drugs in nanoparticles, provides a small boost in performance that is not worth the effort, time, and development costs that have to be invested in the pharmaceutical industry. Moreover, the added complexity of nanoparticles also discourages their use to deliver drugs. Thus, drugs that could improve their effectiveness when administered by nanocarriers are not formulated in this way [5].

## 2. NANOTHERAPY FOR CANCER: SMART NANOPARTICLES AS DRUG-DELIVERY SYSTEMS

Cancer is a major public-health problem worldwide. More than 30% of all people will develop some form of cancer during their lifetime. For instance, it is the second lead- ing cause of death in the United States, and is expected to surpass heart diseases as the leading cause of death in the next few years [7].

According to the estimation of the most common cancers expected to occur in the United States in 2015, most com- mon cancers among men will be prostate (about one-quarter of new diagnoses cancers), lung and bronchus, and colorectal cancers. The most commonly diagnosed cancers in women will be breast (expected to account for 29% of all new cancers), lung and bronchus, and colorectal [8]. (Fig. 2) shows the estimated data of predicted deaths by cancer in the United States for five leading cancer types by gender.

Cancer is a complex multistep process where cells attain certain hallmark properties because of both genetic and epi- genetic alterations [9]. Diagnosis is undoubtedly the first and most important step in cancer therapy. The first therapeutic option for patients with solid tumors is usually surgery, to remove cancer cells. Afterwards, and to destroy any possible remnants of the tumor, the area should be irradiated. Simultaneously or in a later phase, chemotherapy can be used to kill residual can- cer cells and possible metastases [10]. Conventional anticancer therapies are distributed non-specifically in the body and damage both cancer and healthy cells in a state of division. This involves the use of suboptimal treatment to prevent excessive toxicities [11]. Furthermore, many effective drugs are hydrophobic or, if soluble, never reach their destination. Moreover, the multiple and interconnected pathways of carcinogenesis complicate efforts for providing effective therapies. Therefore, a paradigm change is underway as researchers are working to design therapies for specific cancer phenotypes [12].

While the search for new strategies for cancer therapy continues, it becomes increasingly evident that targeted nanovehicles, designed to achieve the specific organ is the key to formulate effective treatment. The greatest advantage of the application of nanomedicines in cancer lies in its potential to create novel structures with enhanced abilities to translocate through cell membranes, thereby, enhancing their delivery efficiency. The benefits of developing nanoparticles as drug-delivery systems include enhancement of pharmacological activity, delivery of more than one therapeutic agent for combination therapy, solubility —as many drug molecules can be incorporated in the particle matrix— stabil- ity and bioavailability (designing nanoparticles with optimal size and surface properties to increase their circulation time in the bloodstream) [13]. These nanoparticles could also im- prove the protection from toxicity and physical and chemical degradation, sustained delivery [14], feasibility of variable routes of administration [15], facilitation of the drug trans- port across critical and specific barriers [16], etc. Engineer- ing materials on this scale allow for novel medical therapies, which are able to release the associated drug to the target tissue in a controlled manner, improving the specificity and resulting in decreased side effects for patients [13]. Another objective is to overcome the development of multiple drug- resistance mechanisms that make this treatment ineffective in a high percentage of cancer cases because cancer cells can evade the cytotoxicity [17, 18]. P-glycoprotein is the best known and most extensively investigated mechanism of drug resistance [19]. Upon entering in the cell, nanoparticles are involved in an endosome allowing them to accumulate in the cells without being recognized by the P-glycoprotein efflux pump, resulting in the increased intracellular concentration of drugs [20]. Moreover, receptor-targeting ligands strategies that are usually internalized via receptor-mediated endocyto- sis, may have particular potential for overcoming drug resistance. Thus, a folate receptor–targeted, transferrin- conjugated paclitaxel-loaded nanoparticles [21] and doxorubicin loaded pH-sensitive polymeric micelles [22] showed greater inhibitory activity against drug-resistant MCF-7 cells and/or xenografts than did their non targeted drug-free counterparts.

Anticancer nanomedicines require the integration of drugs into smart nanoparticles. Diverse platforms can be used to combine different drugs into a single nanotherapeutic agent for synergistic therapeutic benefits. These platforms are submicron-sized particles (10 – 1000 nm in diameter) [23], devices or systems that can be made using a variety of materials (including polymers, lipids, viruses, and organometallic compounds, among others) [20]. Moreover, all nanoparticles based on drug-delivery vectors must be able to both transport and release the drug to a specific location. They can be classified by their physical form or functional properties, both of which should be adapted to the specific needs of the drug to be delivered, and the intended therapeutic use [24].

Although nanomedicine is a relatively new field of science, many new smart nanoparticles have been developed over the past three decades for drug delivery. These plat- forms are generally classified into the following categories: polymer-based drug carriers, lipid-based drug carriers, viral nanoparticles, carbon nanotubes, ceramic nanoparticles, and metal-based nanoparticles. (Fig. 3) shows some examples of drug-delivery platforms.

### 2.1. Polymer-based Drug Carriers

A plethora of different polymeric molecules have been investigated to generate therapeutic carriers. They include polymer-protein conjugates, drug-polymer conjugates and supramolecular drug-delivery systems. Among the many polymers that have been proposed, only a few have been accepted into clinical practice. Specifically, biodegradable polymers are highly preferred due to outstanding bioavailability, better encapsulation, control release, and less toxic properties.

The use of polymeric substances offers various advantages. One of the most important is the incorporation of hydrophobic drugs to some polymer chains that enhance water- solubility, changing the biodistribution pattern [25]. Another advantage could be found using mucoadhesive materials, such as chitosan, since it increases intimacy of contact be- tween the drug-containing polymer and cell membranes [26]. In the field of cancer therapy, site-specific drug delivery nanocarriers are sought, locating in the external part of the nanoparticles different molecules that favor cell- internalization by a reception mediation [27]. More recently, great attention is being paid to the development of novel multifunctional nanoparticles, composed of different mole- cules that are able to co-deliver multiple components, target the delivery of drugs to the specific tumor cells, and release the therapeutic agent [28]. Several types of polymer-based drug carriers have been tested as potential drug-delivery systems for cancer treatments, including polymeric nanoparticles, dendrimers, hydrogels and micelles. Below a summary of these types is presented.

### 2.1.1. Polymeric Nanoparticles

Polymeric nanoparticles are engineered from biodegradable and biocompatible polymers and they have been synthesized using various methods based on the application needs and the type of drugs to be encapsulated [29]. These particles are solid-matrix systems, in which the drug is dispersed within the particle or conjugated to the polymeric backbone. Polymeric nanoparticles have been developed to encapsulate either macromolecules such as nucleic acids and proteins as well as hydrophobic or hydrophilic small drug molecules. In an effort to avoid their recognition and removal by the mononuclear phagocyte system and thereby achieve long circulation time, nanoparticles are usually coated by other protective polymers, such as poly(ethylene glycol), poly(ethylene oxide), or poly(acrylic acid), or dextran. How- ever, this layer of protection should not be very dense be- cause the cellular uptake would also be hindered [30].

Different structures of polymeric nanoparticles are used in cancer treatment, as very recently reviewed by Wang et al. [31]: linear polymers, star polymers, cylindrical polymerbrushes, macrocyclic polymers, macrocyclic brushes, and hyperbranched polymers.

Natural polymers (i.e. chitosan, heparin, albumin, etc.) have been used in FDA-approved drug-delivery formulations. As an example, low-molecular chitosan nanoparticles decorated with a hydrophilic shell of methoxy polyethylene glycol and a core of cholesterol (to improve drug solubility) showed good efficacy of encapsulation and delivery of paclitaxel [32]. In addition, nanoparticles formed by silk fibroin (a natural protein polymer) have stirred great excitement in the field of drug delivery, due to their ability to deliver anti- cancer drugs [33]. Nevertheless, the heterogeneity of natural polymers, cost, and manipulation difficulty, make it much more usual to develop synthetic polymers with enhanced drug-delivery potential.

Numerous nanoparticles are in pre-clinical or clinical development [34]. For instance, nanoparticles comprising hydrophobic copolymers —such as poly (lactic-co-glycolic acid) (PLGA) and poly(alkyl cyanoacrylate) (PACA)— have been used to co-encapsulate chemotherapeutic agents and inhibitors of the Multidrug Resistance (MDR) mechanism for delivery to various types of cancer [35]. Another example can be found in nanoparticles made of a core of polycaprolactone and a hydrophilic shell of polyethylene glycol, that have shown their potential usefulness in anti-tumor treatment encapsulating and delivering cisplatin [36].

2.1.2. Hydrogels

Hydrogels are hydrophilic polymer networks with vari- able structures. They can be classified into those that have a linear structure, and those exhibiting polymeric network with three-dimensional covalent crosslinking. Covalent bonds between chains affect the overall properties of the hydrogel (i.e. swelling ability, elasticity, and drug loading capacity). They are able to absorb from 10-20% (an arbitrary lower limit) up to thousands of times their dry weight in water, a property attributed to the presence of hydrophilic groups in their structure [37]. This makes them very useful as potential drug carriers in the form of micro- or nano-particles. In addition, some hydrogels possess features of fluid transport and stimulus responsive characteristics (e.g. pH, temperature, salinity, or light) [38].

A number of synthetic hydrogels have been studied for anticancer purposes. Some have been used to model micro-environments of tumor vascularization [39], and many others for drug delivery [40]. For example, Cheng et al. prepared thermosensitive hydrogels based on poly (y-ethyl-l- glutamate)-poly(ethylene glycol)-poly(y-ethyl-l-glutamate) triblock copolymers (PELG-PEG-PELG) for localized and sustained delivery of paclitaxel (PTX) in vivo. The results demonstrated that the PTX-incorporated hydrogels could efficiently suppress the tumor proliferation, and did not result in obvious side effect [41]. Recently, Seo et al. have successfully prepared injectable intratumoral hydrogels composed of a diblock copolymer based on poly(ethylene glycol and polycaprolactone (MPEG-b-(PCL-ran-PLLA)) containing 5-fluorouracil. A single injection of this system (applied locally in the tumor) resulted in a significant tumor growth inhibition [42].

2.1.3. Dendrimers

Dendrimers are synthetic polymers with highly branched structures consisting of an initiator core and multiple layers with active terminal groups. Each layer is called a generation (the core is denoted as generation zero) and they are formed of repeating units [43]. Due to their specific molecular structure and the possibility of incorporating a large number of functional groups at spatially defined positions, dendrimers are ideal candidates to develop smart nanocarriers for bio- medical applications. They have been used as antimicrobial and antiviral drugs, in tissue engineering, drug delivery, gene transfection, and other capacities [44-49]. Another advantage lies in the fact that bioactive agents may be encapsulated either into the interior of the dendrimers or can be chemi- cally attached or physically adsorbed onto the dendrimer surface [50].

In this context, Ren et al. developed a poly (amidoamine) (PAMAM) dendrimer that simultaneously delivers genetic material and a chemotherapeutic agent. That is, 5- fluorouracil (5-FU) was encapsulated in the cavities of the dendrimer core via hydrogen bonds, while an antisense microRNA (miR-21) was complexed to the surface [51]. The synchronic delivery of the two therapeutic agents resulted in synergistic anticancer efficacy, apoptotic activity, and de- creased migration ability of the cancer cells compared to each agent alone. Zhang and Shi coated G5-PAMAM- dendrimers with folic acid and doxorubicin, finding a multi- functional system that might serve as a valuable nanodevice for targeting cancer chemotherapy [52]. Finally, Kaminskas et al. explored the utility of a PEGylated polylysine dendrimer conjugated to doxorubicin to promote the controlled and prolonged exposure of lung-resident cancers to cytotoxic drug. They demonstrated >95% reduction in lung-tumor bur- den in rats after two weeks of treatment [53]. This success suggests that PEGylated dendrimers have the potential as inhalable drug-delivery nanosystems against lung cancers.

2.1.4. Micelles

Polymeric micelles are formulated through a self- assembly process using block copolymers consisting of two or more polymer chains with different hydrophilic character. These copolymers spontaneously assemble into a core-shell structure in an aqueous environment. Hydrophobic blocks form the core where any hydrophobic drug can be carried while the hydrophilic blocks form the shell. Polymeric micelles are usually more stable in blood than are liposomes and other surfactant micelles because the amphiphilic co- polymers usually employed have a considerably lower critical micelle concentration value. Due to such unique proper- ties, polymeric micelles have been well demonstrated to be effective drug carriers for cancer therapy and image-based analysis of tumors [54-56]. In addition, it is currently possible to confer pH-sensitive properties to some polymeric micelles, improving the drug-release process in anticancer treatments [57, 58].

Wang et al. showed the specific binding of paclitaxel- loaded micelles targeted with a MCF-7 cell-specific phage protein [59]. Ke et al. designed micelles loaded with both thioridazine —which has been reported to kill cancer stem cells— and doxorrobicin, providing a promising strategy for breast-cancer treatment by targeting both cancer and cancer stem cells with this combination therapy [60]. Yang et al. synthesized mPEG-b-PDLLA micelles loaded with docetaxel and demonstrated clear metastasis inhibition of 4Ta murine cells in the lung of mice. Micelles were capable of internalizing in the 4T1 cells via endocytosis, significantly decreasing tumor metastasis [61]. Finally, Hu and collaborators have recently shown a complete regression of breast tumor using a crossed-linked thermosensitive polymeric micelle linked to docetaxel by a covalent conjugation via hydrolysable ester bonds [62]. Docetaxel-micelle complexes showed great in vivo efficacy in reducing and establishing MDA-MB-231 tumors in mice, even after a single intravenous injection. The preclinical data shown by these authors would support clinical translation of this novel smart nanosystem for cancer treatment.

## 2.2. Virus-based Nanoparticles

Viruses can be considered living nanoparticles with a core–shell structure. The infectious agents are in the core surrounded by a shell comprised of proteins or proteins embedded in lipid membranes. They range in sizes from 10 nm to over a micron and can be found in a variety of distinctive shapes (most abundantly icosahedrons, spheres, and tubes). They can be either isolated directly after inducing the expression of the coat-proteins in eukaryotic cells, or assembled in vitro from coat-protein subunits purified from any recombinant host, from mammalian cells to bacteria [63]. The main advantage of virus-like particles is their morpho- logical uniformity, biocompatibility, and easy functionalization. They have been extensively used for delivery of different substances, vaccines, and gene therapy due to their high gene-transfection efficiency and fusogenic cell receptor- binding properties. However, their implementation into clinical therapy has been plagued with significant draw- backs.

For different substances to be encapsulated in virus-like nanoparticles, it is usual to resort to simple supramolecular self-assembly and disassembly processes. The specialized literature indicates that the functional cargoes usually packaged with these kinds of assembly processes are genetic material or other supramolecular structures (such as liposomes or micelles) that are mixed with DNA. This is why the ex- tension of these virus-like nanoparticles for the "encapsulation" of anti-tumor drugs can be envisaged, but it is rarely used today. Nevertheless, the viral capsid, once it is formed, can be used to deliver drugs if the capsid is loaded with the chemotherapeutic molecule through covalent attachment. In this case, the presence of reactive groups in the protein cage is mandatory. With this procedure based on chemical attachment, virus-like nanoparticles have been loaded with doxorubicin by Aljabali et al. [64], and Zhao et al. [65].

Actually, the use of viruses for anti-cancer treatments is not focused on their potential drug-delivery ability. Rather, its use is based mainly on a dual ability that viruses possess: the stimulation of the immune system of the patient, and the direct cell lysis they provoke upon infesting a tumoral tissue. Oncolytic viruses are engineered to self-amplify and replicate selectively within cancer cells and induce toxic effects such as cell lysis and apoptosis, while leaving normal non- transformed cells intact (or nearly so) [66]. Steadily more results show that their targeted infection of the tumor has the potential to create an "inflammatory storm" that arouses the innate and adaptive immune responses against tumors [67]. Please see the recent and detailed information on this topic in references [68-70]. As shown in the specialized literature, the vaccinia virus family is the most promising one to be used in cancer therapy.

Additionally virus-like nanoparticles have a growing application in the field of imaging of tumors, although this topic will not be covered in this review.

## 2.3. Carbon Nanotubes

Carbon nanotubes (CNTs) are allotropes of carbon with a cylindrical nanostructure. Nanotubes have been constructed with length-to-diameter ratios of up to 28,000,000:1 [71]. CNT can be imaginatively produced by rolling up a single layer of graphene sheet (single-walled CNT; SWNT) or by rolling up many layers to form concentric cylinders (multi- walled CNT; MWNT) [72].

Although CNTs have several unique properties related to their stiffness-flexibility and electric conductivity, non-functionalized (pristine) CNTs are poorly soluble and highly cytotoxic. A great variety of safety concerns of CNT can be found in references [73, 74]. In summary, the toxicological profile in the human body may exert an effect at the cellular level, tissue level, or whole-body level, and it has been demonstrated that the toxicity depends on: i) the length of the CNT used (the longer, the greater the bio-incompatibility), ii) the surface hydrophicity (since a surface chemical functionalization of the CNTs with small hydrophilic groups en- hances their biocompatibility), and iii) the CNT individualization (i.e. a lack of any aggregation process), since the higher the degree of individualization, the higher the degree of urinary excretion.

As mentioned above, if the CNT surface is functionalized with soluble and biocompatible materials, its solubility in water increases as does s its biocompatibility. Functionalized carbon nanotubes have been successfully investigated for several biomedical applications such as proteins, nucleic acids, and drug carriers [75-77]. The potential of CNTs for delivery of anticancer agents might be attributed to their exclusive physicochemical properties, especially their ability to cross various biological barriers by a piercing (needle-like) mechanism. Application of CNTs for the targeted delivery of drugs has become one of the main areas of interest for different research groups. In this context, in vitro delivery of paclitaxel attached to single-walled CNTs showed higher efficacy in inhibiting tumor growth than delivery of paclitaxel alone [78]. Furthermore, Liu et al. demonstrated that the control release of the chemotherapeutic drug 7-Ethyl-10-hydroxy- camptothecin (SN38) by SWCNTs conjugated with antibody C225, can achieve targeted therapy against EGFR over- expressed colorectal cancer cells, suggesting that SWCNT could be a good chance to form carriers for targeting therapy [79].

## 2.4. Lipid-based Drug Carriers

### 2.4.1. Liposomes

Liposomes are spherical lipid vesicles in which the aqueous phase is encapsulated by a bilayer (or more) of amphiphilic lipid molecules of synthetic or natural origin [80]. Liposomes has become one of the first nanoparticulate plat- forms for drug delivery because of their unique characteristics: (a) biocompatibility and biologically inert profiles with- out antigenic or toxic reactions in most patients; (b) a long life in blood enhanced with the pegylation of the surface (stealth liposomes); and (c) the easy control of physico-chemical properties as size or surface characteristics which provide multi-functionalized liposome systems [81] coated with polymers and different ligands to target specific cells [82]. Immunoliposomes, with monoclonal antibodies or anti- body fragments attached to the surface of liposomes, represent the better part of active targeting studies [83]. The different structural parts of the liposome, that is, the internal water, the phospholipid bilayer or the bilayer interface, can be used to carry both, hydrophilic and hydrophobic therapeutic agents. For each particulate case the amount of drug loading is different and must be carefully optimized to develop the different formulations [83].

Taking into account the "first-generation" of drug nanocarriers, liposomal drug delivery systems are the most used in clinical applications and with the better result. In this con- text, Doxil®, a pegylated liposomal formulation with doxorubicin was the first nanocarrier of liposomal characteristics approved in 1995 by the USFDA (Food and Drug Administration of USA) indicated for the Kaposi's sarcoma related with AIDS [84]. Moreover, in 2012, Marqibo®, a liposome composed of egg sphingomyelin and cholesterol was used to battle acute lymphoblastic leukemia [85]. Be- sides these two compounds, only around 10 liposomal formulations have been approved for clinical use, half of these to deliver anticancer drugs on the basis of reduced side effects [83, 86]. However, most of them are in clinical and preclinical trials [81], preferentially focused for the therapy of different types of cancer and several encapsulating nucleic acids as plasmids or siRNA.

The latest trends have focused on the development of multifunctional liposomes, improving the limited properties of the conventional ones— that is, low stability and targeting properties and limited control of the release profile [81, 87]. A PEGylated version of neutral-lipid-based nanoliposomes has shown promise as highly effective and safe delivery systems for systemic use of siRNA therapeutics [88]. Taking advantage of the subacid environment typical of tumor tis- sues, several pH-responsive nanosystem have been developed for an efficient delivery anticancer drugs inside tumor cells. Phosphatidylethanolamine-based pH-sensitive liposomes surface modified with a cell penetrating peptide (RGD peptide) have been used to enhance the effectiveness of docetaxel treatment, showing good pH sensitivity with higher cytotoxicity and cellular uptake ratios [89]. The same pH-sensitive property system has been achieved by implanting octylamine-graft-poly aspartic acid onto liposomes. The structure of grafted systems bolsters the stability of the drug carrier and raises tumor-cell toxicity due to the fast release of antitumor drug cytarabine, subject to the octylamine in an acidic environment [90]. Temperature-sensitive liposomes have also been proposed. A recent example consists of liposomes with a triple functionalization: PEGylation, ther- mosensitive chains and target specificity by the surface conjugation of antibody Herceptin. This system retains a hydro- phobic cytotoxic drug, Docetaxel, inside but immediately released at temperatures above 40ºC [91].

2.4.2. Solid-lipid Nanoparticles

Solid-lipid nanoparticles have been developed by Müller et al. and Gasco et al. at the beginning of the 90s [92]. The solid-lipid nanoparticles (SLN), which range in size between 40 and 1000 nm, are composed of a biocompatible/biodegradable matrix of solid lipids stabilized by surfactants [93]. This lipid core maintain their solid state at room and body temperatures SLN present several advantages for their use as drug nanocarriers: the effective and economic production process with a low cost of the solid lipid com- pared to phospholipids and the absence of organic solvent; the possibility of large-scale production; good tolerability and biodegradability; physical stability; amenability to encapsulating hydrophilic and hydrophobic bioactive mole- cules or drugs; and the potential application in alternatives routes of administration instead of parenteral ones [92, 94-96]. However, SLN has certain disadvantages, mainly the low drug loading, especially with peptides/proteins and other hydrophilic molecules; and the drug expulsion from the nanoparticle during storage time as a consequence of trans- formations of the solid lipidic components [87, 96]. A second generation of SLN has been developed to minimize these drawbacks,. Nano-structured Lipid Carriers (NLC) are based on a mixture of solid and liquid lipids, which reduces the homogeneity of the matrix structure and increases the solubility for drugs [87].

Various anticancer drugs, including etoposide [97], methotrexate [98], idarubicin [99], paclitaxel and docetaxel [100], or edelfosine [101] have been incorporated into SLN and evaluated with positive and promising results. overcoming many troubles observed in the conventional therapies. Due to the low tumor uptake shown by SLN and NLC using the intravenous administration route, the interest in both as alternative to other nanocarriers, lies in its application through other routes of administration, such as pulmonary, oral, ocular, and especially dermal ones, making them the subject of interest for numerous studies [102].

Several taxanes have been directly delivered into the lungs by inhalation using multifunctional NCL. These nano- carriers have been coupled to siRNA, preventing drug resistance, and targeted with a synthetic hormone specific to the recognition of overexpressed receptors of lung-cancer cells [103]. The molecular mechanism of SLN and nucleic acid assembling process is under study in order to gain in-depth knowledge of this complexation process and how the nucleic acid carried may be released intracellularly [104]. SLN have also shown a protective effect of reducing the gastrointestinal toxicity of an antineoplastic drug, such as edelfosine. This situation has been tested in in vivo experiments by oral administration in mice [105]. As in other nanocarriers, hydrophilic polymers or surfactant coatings improve the stability properties of the SLN and NCL. Chitosan-coated SLN represent a promising system for the oral drug delivery of hydrophobic drugs [105] and pegylated NLC encapsulating Amoitone-B was found to improve the cytotoxic effect of this drug in human-colon and liver-cancer cells in a dose- and time-dependent manner due to the solubility of PEG in the cell membranes [106].

2.4.3. Lipid Nanoemulsions

The nanoemulsions based in lipids are oil/water colloidal dispersions in which the nanoparticles (dispersed phase) are stabilized in water (dispersed phase) by surfactants and co- surfactants self-organized at the oil/water interface [107]. Emulsions are usually classified into two main categories: microemulsions and nanoemulsions. These latter are emulsions with a mean droplet diameter on the nanometer scale and generally composed of a dispersed oil phase within a continuous aqueous phase having a radius of less than 1000 nm, although some authors consider 500 nm as the upper limit [107, 108], while other authors place the cutoff at 200 nm [109] or even 100 nm [110]. However, it is generally accepted that droplet diameters in nanoemulsions are less than ~200–300 nm in size, and thermodynamically it is a non-equilibrium state which needs to be stabilized by altera- tion of the attraction–repulsion potential, for example, by the inclusion of stabilizing surfactants [111]. Nevertheless, the kinetics of destabilization is so slow that they can be considered stable from this point of view, in a different way to mi- croemulsions, which can reach up to 1 μm in size as they are thermodynamically stable [112-114]. Nanoemulsions are also usually referred to as lipid nanocapsules, because morphologically they are formed by an oily core encapsulated inside a spherical polymeric shell (thick steric barrier at the droplet interface) [113]. However, some authors reserve nanocapsule nomenclature for structures formed by a more rigid polymeric shell prepared by interfacial polymerization or interfacial nano-deposition [115, 116]. In any case, both lipid nanocapsules and lipid nanoemulsions show simultane- ously characteristics of lipid core micelles and polymer nanocapsules with a shell composed by a blend of phospholipids and surfactants [117].

Emulsion-solvent evaporation is a widely used technique to encapsulate lipophilic drugs inside lipid nanocapsules or nanoemulsions, constituting a simple oil-in-water (O/W) emulsion. A handicap of these nanosystems is the reduced loading level of biomolecules and drugs of hydrophilic character due to their leakage to the aqueous phase before the solidification process of the polymeric surface shell take place. A modified water-in-oil-in-water (W/O/W) double emulsion allows to reduce the loos of active molecules pre- serving at the same time as the biological activity of the biomolecules in the aqueous phase [118, 119]. Vegetable oils (fatty acyls) or middle-chain triacylglycerols are the typical compounds of the dispersed O-phase in most nanoemulsions, while the shell can be formed by natural phospholipids, alcohols and a wide variety of polymers/surfactants. The use of polymers with hydrophilic segment such as PEG, poly- ethylene oxide, poloxamers, poloxamines, polysorbates or chitosan is widespread [120, 121]. Polymers with these characteristics drive to an increase of the colloidal stability of the nanosystem by means of steric repulsion [122], may help to achieve Stealth® nanosystems that extend the life-time in the bloodstream after intravenous injection by repelling plasma proteins [123].

The use of nanoemulsions as delivery systems is attracting interest in the pharmaceutical industry. Recently, more innovative nanoemulsion formulations have been developed, varying the different components such as the emulsifier and/or the emulsified oil nature [124]. The new functionalities join to the high biocompatibility and biodegradability typical of these emulsions has allowed the access of nanoemulsions to the drug delivery systems market and Etomidat-Lipuro® (encapsulating etomidate) and Diazepam-Lipuro® (encapsulating diazepam) are examples of this commercialization [125]. The main advantages of nanoemulsions in the field of drug-delivery systems for hydrophobic drugs over other nanotechnologies are: inexpensive and easy-to-scale production, low toxicity, independence of dilution, high loading capacity, formulation stability, and reduced side effects [126, 127]. Microemulsions have their main applications in oral or topical delivery [128]. The better stability and small size of nanoemulsions makes them more suitable for parenteral drug delivery and usually, nanoparticles smaller than 100 nm preferentially accumulate in cancer tissues through the so-called enhanced permeability and retention effect [87, 129].

Nanoemulsions have shown a strong potential for drug delivery in cancer treatment and have been widely investigated. For example, pharmacokinetics studies have shown that celecoxib nanoemulsions transdermally applied on rats increased bioavailability as compared to an oral capsule formulation [130]. Orally administered nanoemulsions have also been developed improving the oral bioavailability of hydrophobic drugs such as paclitaxel [131], tamoxifen citrate [132], and raloxifene hydrochloride [133], among others. The enhanced solubility of hydrophobic drugs in the inner oil phase of the emulsion system allows the amount of the drug supplied to be reduced while protecting body fluids and tis- sues from their direct action, thereby minimizing the harmful side effects associated with the intravenously intake of these chemotherapeutic agents [134]. However, although these nanosystems improve pharmacokinetics and bioavailability of lipophilic drugs, they require a higher sophistication by developing multifunctional nanoemulsions with tuneable biological properties [111].

Recent studies on nanoemulsions for parenteral administration have been developed, showing attractive results. For example, lipid nanoemulsions have been used as a platform for the concurrent delivery of the tocotrienol-rich fraction of vitamin E, which previously showed anticancer activity [135]. Promising results have also been shown by the PEGylated nanoemulsions loaded with lycobetaine (LBT)-oleic acid ionic complex against lung cancer, which showed higher LBT levels in blood and longer circulation time, de- creased LBT concentration in the heart, liver, and kidney, higher growth-inhibitory effect, and longer survival com- pared to free LBT [136]. Delivery systems actively targeted to the tumor site or specific tumor cells have been developed, covering the lipid nanocapsule surface with several ligands and biomolecules. Different studies have reported on folate- mediated targeting of antitumor drugs and genes by conjugating folic acid onto polymeric nanoemulsions. In this way, the use of folate-chitosan conjugates covering lipid core- shell nanocapsules has allowed to improve the antitumor activity of these nanosystem [137]. Targeting ligands as monoclonal antibodies allows the development of lipid immuno-nanocapsules in which a specific antibody molecule is conjugated to the nanocapsule surface. Multi-therapeutic systems can be designed using antibodies interfering with signal transduction pathways involved in cancer-cell proliferation [87]. This is the case of lipid immunonanocapsules in which a specific HER2 oncoprotein antiboby was covalently immobilized on a carboxyl-functionalized surface. Docetaxel-loaded HER2 immuno-nanocapsules combine the cytotoxic effect of the loaded taxane and the specific surface antibody, increasing their uptake and the cytotoxicity in HER2 overexpressing tumor-cell lines [138].

2.5. Ceramic Nanoparticles

Ceramic nanoparticles are typically composed of inorganic compounds such as hydroxyapatite, zirconia, titania, alumina and silica [139]. Moreover, metal oxides, and metal sulfides can also be used to form the nanoparticle core. [140]. They are generally bio-inert and have porous structures. These nanoparticles have been proposed as drug- delivery vehicles to carry drugs for various cancer therapies because they can be easily engineered to the desired size, shape, and porosity [141, 142]. Some advantages make ceramic nanoparticles a promising tool for the control drug delivery: the simple conditions required for the preparative processes, high stability, high load capacity, easy incorporation into hydrophobic and hydrophilic systems and different administration routes (oral, inhalation, etc.) [143]. There are no swelling or porosity variations with a change in pH, and they are not vulnerable to microbial attack [144]. They can also be easily functionalized by various molecules due to the presence of negative charge on the surface. It is possible to design a system of an ultra-small size (less than 50 nm) helping them to evade the reticulo-endothelial system (RES) of the body [145].

Titania nanoparticles have been successfully used in in vitro studies, representing a new therapeutic agent for different human cancer cells by means of photoexcitation or ultra- sound processes [143]. The design and synthesis of hydroxyapatite (HA) nanoparticles, the most stable form of calcium phosphate, are focused mainly on bone regeneration. However, other applications have been proposed, for example the use of HA nanoparticles that have shown antiproliferative and proapoptotic on hepatic carcinomas [146]. Mesoporous silica nanoparticles of around 100-130 nanometers of diameter were loaded with camptothecin and targeted with folic acid, and a dramatic improvement was achieved in the potency of tumor suppression (two different human pancreatic cancer xenografts on mice) [147]. However, these systems present a major disadvantage—that is, ceramic nanoparticles are not biodegradable and they might accumulate in the body and have dangerous effects [145, 148].

2.6. Metal-based Nanoparticles

Metal-core-based nanoparticles open the door to the development of multifunctional systems linking drug delivery with both magnetic response and imaging abilities that involve the concept of theranostic systems, combining simultaneously therapy and diagnostic [149]. Metal nanoparticles can be synthesized in extremely small sizes (around 50 nm) and thus the large surface area provides the ability to carry a relatively higher dose of drugs. One of the most commonly used is gold nanoparticles (AuNPs) because they are bio- compatible, easy to synthesize, characterize, and surface modify. The presence of surface plasmon resonance (SPR) bands is responsible for their large absorption and scattering cross-sections, which are 4 to 5 orders of magnitude larger than that of conventional dyes [150]. Thus, gold nanoparticles are new agents that are being evaluated for biological sensing, drug delivery, and cancer treatment [151]. Recently, it has been shown that AuNPs inhibit the proliferation of cancer cells by abrogating MAPK-signaling and reverse epithelial-mesenchymal transition (EMT) in cancer cells by reducing the secretion of a number of proteins involved in EMT. Therefore, an inhibition of tumor growth and metastasis in two separate orthotopic models of ovarian cancer were observed [152]. Additionally, potential applications of AuNPs have recently been studied and administrated in phase I & II clinical trials for cancer treatment [153].

Other typical metal nanoparticles are core-shell magnetic nanoparticles. They consist of a metal or metal oxide core coated with inorganic or polymeric shell to improve their stability and render the particle biocompatible [154]. Metal nanoparticles can be used as thermal-release triggers when irradiated with infrared light or excited by an alternating magnetic field. Higher cell inactivation was found when gamma radiation (current radiotherapy) interacts with high- Z-element nanoparticles (photoelectric effect), resulting in a high-level electron release in the media of the radio-resistant cells. Superparamagnetic zinc ferrite spinel ($ZnFe_2O_4$) nanoparticles are an example of this situation [155]. In addition, in magnetic drug targeting, magnetic carrier particles with surface-bound drugs are injected into the vascular systems and then captured at the tumor via a locally applied magnetic field. Several studies have demonstrated the application of magnetic nanoparticles for drug delivery. Chen et al. developed a magnetic drug-delivery system that may find potential applications in cancer treatment, in which doxorubicin chemically bonded to $Fe_3O_4$ nanoparticles, were embedded in a polyethylene glycol functionalized porous silica shell [156]. Another group used external magnets to demonstrate that gold-coated iron-oxide nanoparticles with the active component of cisplatin can be accumulated in specific regions, and that cell-growth inhibition was localized in those areas [157]. $Fe_3O_4$ nanoparticles with magnetic properties were modified with chitosan crosslinked carboxymethyl- $\beta$-cyclodextrin enhanced the drug-loading capacity. The hydrophobic anticancer drug 5-fluorouracil was satisfactorily encapsulated and its release behavior was pH dependent and diffusion and swelling controlled [158].

Currently, many other metal-based nanoparticles, as well as various hybrid surface-functionalized nanoparticles, are being extensively used for drug-delivery applications, such as platinum [159], silver, and palladium nanoparticles [160, 161]. Quantum dots, fluorescent semiconductor nanocrystals, have also been extensively investigated for drug delivery and imaging [162]. They combine small size, versatile surface chemistry and outstanding optical properties [163]. The anti-cancer drug epirubicin was conjugated with iron oxide nanoparticles for delivery and image function in a tumor-bearing rat-brain model. In this case, drug delivery was con-ducted by magnetic targeting with promising results [164]. Docetaxel delivery and cell imaging have been combined in lanthanide oxide nanoparticles. This terasnostic system has a cytotoxic effect against human cervical carcinoma cells, and the function of optical imaging thanks to the intrinsic luminescence of nanoparticles [165]. PLGA nanoparticles encap-sulating simultaneously an anticancer drug, busulfan, with inorganic agents such as iron oxide nanoparticles or manganese-doped zinc sulfide simultaneously allow delivery of the anticancer drug and knowledge of the cell uptake and biodistribution by in vitro and in vivo imaging [166].

Nanoscale metal organic frameworks (NMOFs) are hybrid materials formed by the self-assembly of bridging ligands and metal-connecting points. Preliminary biomedical applications of NMOFs have focused on their use as delivery vehicles for imaging contrast agents and molecular therapeutics. Several different therapeutic molecules were loaded within porous iron-carboxylate NMOFs at unprecedented levels. The NMOFs showed sustained drug release with no burst effect, and in vitro assays revealed that the nano-encapsulated drug showed efficacy similar to that of the free drug [167]. In addition to their use as drug-delivery systems, metal nanoparticles can act directly as pharmaceutically active agents. This recent idea explores the therapeutic application of inorganic nanomaterials and the development of new anticancer agents based on metal nanomaterials. Several in vitro and in vivo reports suggest that certain metal-based nanoparticles can interfere with the antioxidant cell-defense mechanism, leading to an increase in the reactive oxygen species, thus being effective to induce apoptosis and auto-phagy in cancer cells at specific concentrations that are not significantly toxic to non-cancerous cells [168].

## 3. DRUG-DELIVERY STRATEGIES

One of the main objectives of nanomedicine is to over-throw the deficiencies of classical chemotherapy, including, firstly, the lack of specificity of the drugs, increasing the probability of adverse side effects caused by the damage of both tumor and normal cells in a state of division. Secondly, the low concentration of active substance at the targeted site, resulting in poor therapeutic effect. Therefore, drugs need to be targeted to the tissue or organ of interest rather than be allowed free systemic circulation.

Nanotechnology allows the possibility of designing novel and more specific therapies that act selectively on the target site. This permits the minimization of adverse side effects since smaller but more effective doses are administered.

### 3.1. Passive Targeting

The accumulation of a drug or drug-carrier system at a specific site is known as passive targeting and may be attributed to physicochemical and pharmacological factors. This type of targeting relies on the disease pathology and the fe tures of tumor tissues, which may favor the accumulation of a drug in target tissues, decreasing non-specificity. It is generally accepted that the vasculature of tumors differs from the surrounding tissue. Compare with normal well-organized vessels, the tumor angiogenesis presents characteristics that improve drug retention, such as extensive vascular density and permeability, imperfect vascular architecture, and im-paired lymph clearance from the interstitial spaces of tumor tissue.

The average pore size of the angiogenic tumor vasculature is estimated to be 100–600 nm, in contrast with the pores of normal blood vessels which are typically <6 nm wide. [169]. This differentiated structure of tumor vessels and this particular tumor vascular permeability, known as "enhanced permeation and retention" (EPR), is important in the delivery of macromolecular anti-cancer agents and nanomedicines (Fig. 4). EPR has shown greater activity and reduced toxicity of many anticancer nanomedicines com-pared with free anticancer drugs [170, 171]. However, there is significant heterogeneity within and between tumor types.

### 3.2. Active Targeting. Potential Targeting Ligands

Paul Ehrlich suggested the concept of drug targeting a century ago and coined the term "magic bullet", referring to an entity consisting of two components. The first one should recognize and bind the target, hence providing precise transport of the drug, while the second one should provide a therapeutic action. Ever since this concept was introduced, many efforts have been made in cancer therapy with the goal of designing targeted therapeutic agents against cancer cells [172]. Active targeting implicates the use of externally con-jugated targeting moieties for enhanced delivery of nanocarriers.

A rapidly growing tumor requires diverse nutrients and vitamins, thus, tumor cells overexpress many tumor specific receptors. To achieve effective tumor-specific drug delivery, nanoparticles are functionalized with targeting ligands that bind to those receptors, which may help internalization upon binding [see (Fig. 5)].

Receptors exploited in targeting drugs to tumor tissues and microenvironments are G-protein- coupled receptors (GPCRs, integrins, folate receptors, transferrin receptors, epidermal growth factor receptor (EGFR), fibroblast growth factors (FGFRs) and sigma receptors. Some less common receptors recently used are follicle- stimulating hormone receptors, C-type lectin receptors, bio- tin receptors, and neuropilin receptors [173].

Other targets that are used to achieve tumor-selective accumulation of drug nanocarriers are those expressed on endothelial cells of tumor vasculature. The angiogenesis de- pendency of tumor growth is a promising target for the development of therapies that control tumor expansion by the inhibition of the growth of new tumor-feeding blood vessels [174-176]. Anti-angiogenesis approaches are effective in limiting tumor growth, with the endothelial cells of the tumor blood vessels as the main targets. Cancer cells are deprived of nutrients and oxygen with the consequent destruction of the tumor [177]. This mechanism remains debated because excessively decreasing vascular permeability would eventually lead to tumor progression (increased tumor hy- poxia) and drug resistant (decreased drug delivery). Thus, due to tumor heterogeneity, angiogenesis strategies are currently proposed to treat non-vascularized areas, improving the delivery of the therapies [178]. The main angiogenic targets, explored by nanomedicine, include the vascular endothelial growth-factor receptors (VEGFRs), $\alpha v \phi 3$ integrins, matrix metalloproteinase receptors (MMPs), and vascular cell-adhesion molecule-1 (VCAM-1) [179].

The types of ligand used to target nanosystems against cancer cells include a wide range of synthetic and natural compounds of different chemical classes. Some of the most widely used are monoclonal antibodies (mAb) and other pro- teins (such as transferrin), nucleic acids (aptamers), small molecules (hyaluronic acid, folic acid), and peptides.

Monoclonal antibodies (mAb) were the first and are still the preferred class of targeting molecules. The first mAb to gain FDA approval for the treatment of cancer was Rituximab in 1997, a chimeric mAb for the treatment of B-cell non-Hodgkin's lymphoma. Trastuzumab, in 1998, a humanized mAb for the treatment of HER2-expressing breast cancer, quickly followed. Bevacizumab was approved in 2004 for treating colorectal cancer and is a tumor angiogenesis inhibitor that binds to vascular endothelial growth factor (VEGF). Same year was approved Cetuximab, which binds to epidermal growth factor receptors (EGFR), for treating colorectal cancer and in 2006 for treating head/neck cancer. , Recent studies have tried to encapsulate chemotherapeutic drugs into nanoparticles and then functionalize the particle surface with mAbs to maintain targeting efficacy [138, 180- 183]. The conjugated antibodies enhance uptake and cyto- toxic potency of the nanoparticles.

Transferrin receptors are overexpressed on cell surfaces when metabolic processes are increased. It has been reported that the membrane transferrin receptor-mediated endocytosis of the complex of transferrin-bound iron and transferrin receptor is the major route of cellular iron uptake via clathrin-coated pits, with subsequent traffic to endosomal compartments. This uptake pathway has been successfully exploited for the delivery of anti-tumor drugs, proteins, and therapeutic genes [184-187].

Aptamers are RNA or DNA oligonucleotides capable of binding to target antigens with specificity analogous to anti- bodies thanks to their unique three-dimensional conformations originated by intramolecular interactions. They have many favorable characteristics including small size (20 to 80 bases), lack of immunogenicity, and ease of isolation [188]. The most important success of aptamers so far has been the development of FDA-approved aptamers that are able to bind VEGF, a protein involved in angiogenesis [189]. The conjugation of aptamers to drug-delivery nanoparticles resulted in improved targeting and more efficient therapeutics, as well as more selective diagnostics [190].

Hyaluronic acid (or hyaluronan) (HA) is a linear, negatively charged and natural polysaccharide found in the extracellular matrix and synovial fluids of the body. It is responsible for various functions within the extracellular matrix such as cell growth, differentiation, and migration. It has been shown that the HA level is elevated in various cancer cells. The higher concentration of HA in cancer cells is believed to form a less dense matrix, thus enhancing cell's motility as well as its ability to invade other tissues. Hyaluronic acid has been investigated as a targeting moiety of the drug conjugates or nanoparticles for cancer therapy, because it can specifically bind to various cancer cells that over-express CD44, an HA receptor [191-193]. Folic acid belongs to the vitamin B family but it is not produced endogenously by mammalian cells. It is important in the formation of new cells because is an essential precursor for the synthesis of nucleic acids and some amino acids. Folate requires internalization by cells via either receptor- mediated endocytosis or a carrier-based uptake mechanism. There are two membrane-bound folic acid receptors, FR-a and FR-p. FR-a is overexpressed in a large number of human cancer tumors including malignancies of the ovary, brain, kidney, breast, and lung and it has a high affinity for folic acid, which results in high uptake by FR-positive cells [194]. Folic acid binds to tumor cells 20-times more than to normal epithelial cells or fibroblasts. Because of these attractive characteristics, folate conjugation has become a widely used strategy for targeting drug-delivery systems [137, 195-198].

The functionalization of nanoparticles with peptides presents the advantage that high-affinity sequences can be dis- covered through screening of combinatorial libraries. Cell permeating and fusogenic peptides from pathogens or toxins and peptides, randomly derived from technologies such as phagedisplay, are also commonly used as targeting moieties [199]. Among single nuclear localization (NLS) peptides, trans-activating transcriptional activator (TAT) peptide has been shown to be an efficient molecule for translocating nanoparticles into cell nuclei via the binding of import receptors, importin a and p. In a recent publication a peptide has been used to conjugate onto mesoporous silica nanoparticles for nuclear-targeted drug delivery of doxorubicin for the first time [200].

Lipid nanoparticles targeted with peptides are among the most often studied [201].

## 4. INTRAVENOUSLY ADMINISTERED NANOPARTICLES

For parenterally administered drugs, Interactions with blood components, kinetics and systemic distribution are very important for parenterally administered drugs. In a intravenous treatment, therapeutic agents will find biological barriers such as hepatic and renal clearance, enzymolysis and hydrolysis, cellular uptake, endosomal/lysosomal degradation, and spleen filtration. In the case of anticancer drugs, the efficiency is also influenced by their poor solubility, low stability, and high toxicity for normal cells.

Nanomaterial carriers can improve the biodistribution and prolonged blood circulation of therapeutics, thereby sig- nificantly increasing drug efficacy, and decreasing the dos- age. When nanoparticles are parenteral administered the in vivo fate of a drug is determined by the type of the drug- delivery system, not by the drug properties. Examples of different nanostructured carrier systems and the properties that influence their suitability for parenteral administration are shown in (Table 1).

The filtration of blood through the kidneys effectively removes nanoparticles with diameters smaller than 10-20 nm, and the filtration through inter-endothelial slits in walls of the splenic sinus removes particles larger than 200 nm. Liver fenestrations found in the sinusoids also temporarily remove nanoparticles from the circulation. These blood filters suggest that the size of nanoparticles should be no smaller than 20 nm and no larger than 200 nm if long circulation within the blood is desired. The mononuclear phagocyte system (MPS), comprised of phagocytic cells, primarily monocytes and macrophages, plays a vital role in clearing the nanoparticles. MPS has large populations residing in the spleen, liver, and lymph nodes. The recognition of nanoparticles by these cells is greatly enhanced following a process called opsonization. This process is the adsorption of protein entities capable of inter- acting with specific plasma-membrane receptors on monocytes and various subsets of tissue macrophages, thus, promoting particle recognition by these cells [211]. On the other hand, the absence of opsonins in the adsorption pattern and the presence of dysopsonins lead to the circulation of particles in the bloodstream. Properties such as nanoparticle size, surface charge, hydrophobicity/hydrophilicity, elasticity [212, 213] and the steric effects of particle coating can dictate nanoparticle compatibility with the immune system [214]. Hajipour et al. have recently shown that plasma samples from humans with different diseases influence the deco- ration of the proteins that adsorb onto the nanoparticles, introducing the pioneering concept of "personalized protein corona" [215].

### 4.1. Biomolecular Corona: What Cells See of Particles?

When nanosystems are in a physiological environment, they rapidly adsorb biomolecules such as proteins and lipids onto their surface, forming a biomolecular "corona" (Fig. 6), an interface organization that may be loosely divided into two components termed the "hard" and "soft" coronas, with (respectively) "long" and "short" typical exchange times. The corona will change composition if the nanoparticle moves to another compartment or fluid. Tenzer et al. found that the corona formed rapidly after the particles had been mixed with plasma and it could be detected even at the first measurement point (at 30 s) [216].

This corona surrounding the particle changes its original surface charge and chemistry, size, solubility, aggregation and, hence, it changes the dialogue of nanoparticles with the cells, influencing the trafficking, in vivo biodistribution and cellular uptake because these proteins adsorbed onto the sur- face interact with cells through receptors on cell plasma membrane [217]. The protein corona influences macrophage capture since proteins involved in the process of opsonization (opsonins) can be adsorbed onto the nanoparticle surface. Macrophage scavenger receptors are a broad group of phagocytic receptors that are responsible for the elimination of blood-borne viruses, pathogens, and negatively charged ligands. For ex- ample, opsonins such as IgG, complement factors, and fi- brinogen promote the phagocytosis, removal of nanoparticles from the bloodstream and concentration in the liver and spleen, while dyopsonins such as albumin and apolipoprotein help the longer circulation of nanoparticles in the body [94]. Human serum albumin, when adsorbed onto the surface of polystyrene microparticles, was reported to inhibit their phagocytosis by dendritic cells [218].

For many NPs, while removal from the bloodstream is a question of minutes, interaction with cells of distant organs may be relevant hours or days after exposure. Nanoparticles functionalized with hydrophilic polymers such as polyethylene glycol (PEG) and poloxamers show improved circulation lifetime properties and decreased macrophage recognition of many types of nanoparticles by preventing protein absorption [219, 220]. Plasma-protein absorption is influenced in the case of PEGylated nanoparticles by the length and density of the polymer.

It has been claimed that the protein corona can cover/eliminate the targeting moieties on the surface of nanomaterials and, thus, strongly reduces the targeting capability and recognition of the targeting ligand by cellular receptors [221]. Therefore, understanding the interaction of cell plasma membrane and the biomolecular corona around nanoparticles is essential in designing nanoparticle-based targeting drug delivery and treatments.

### 4.2. "Passports" to Avoid Immune-system Clearance

Nanoparticles can be engineered to avoid the immune- system clearance and increase the circulating half-life in the blood. One important reason to search for this is to provide a long-circulating drug reservoir from which the drug can be released into the vascular compartment in a continuous and controlled manner.

A relatively successful approach for prolonging the circulation times of colloidal particles in the blood is to create a steric surface barrier of sufficient density. Poloxamers and poloxamines have been investigated to reduce adsorption of proteins and blood opsonins and thereby increase the half-life of nanosystems. Poloxamers, also known as Pluronic®, and poloxamines or Tetronic®, are amphiphilic nonionic block polymers of hydrophobic propylene oxide (PO) and hydrophilic ethylene oxide (EO). Poloxamers consist of a hydrophobic central polyoxypropylene (POP) molecule, which is flanked on both sides by two hydrophilic chains of polyoxyethylene (POE) yielding structures of the (POE)a (POP)b (POE)a type, as shown in (Fig. 7). Poloxamines are tetrafunctional block copolymers with four POE–POP blocks joined together by a central ethylene diamine bridge [222]. The absorption of these molecules onto the surface of the nanoparticles via their hydrophobic POP fragments provides stability to the particle suspension by a repulsion effect through a steric mechanism of stabilization, because this type of absorption leaves the hydrophilic POE extended outwards from the particle surface in a mobile state. Nanoparticles engineered with poloxamers and poloxamines exhibit reduced adsorption of proteins and blood opsonins [223] and, as a result, they resist ingestion by phagocytic scavenger cells and remain in the systemic circulation for prolonged periods.

Another property presented by poloxamers and poloxamines is the inhibition of multidrug resistance [224, 225]. This phenomenon is the ability of tumor cells to develop resistance to the cytotoxic effects of several chemically unrelated anticancer drugs, associated with the overexpression of proteins such as P-glycoprotein and multidrug-resistance-associated protein [226].

The addition of poly (ethylene glycol) (PEG) has been widely used to increase the circulating half-life of nanoparticles, and it is the preferred method of "masking" nanoparticles from immune recognition. This process is also known as PEGylation [227]. PEG is a linear polyether diol that shows a low degree of immunogenicity and antigenicity [228]. The polymer backbone is basically chemically inert, and the terminal primary hydroxyl groups are available for derivatization. Modification of nanoparticle surface with PEG and can be performed by adsorption, incorporation during the nanoparticle synthesis, or by covalently linking to the surface of nanoparticles [211]. The PEGylation can prevent interactions between the particle surface and the plasma proteins, although adsorption is not completely avoided. Various studies have been conducted to determine how a change in the thickness and density of a PEG coating affects opsonization and biodistribution, showing that the extent of protein ad-sorption depends on the PEG size and grafting density [229] and suggesting conformational transition of PEGs between mushroom and brush states when PEG density increases [230, 231].

Recent progress has shown that it may be possible to dis-guise nanomaterials as "self", avoiding the blood clearance by phagocytic cells, by using a novel peptide sequence mimicking the "marker of self" CD47 protein [232].

## 5. CLINICAL TRIALS AND MARKETED NANOMEDICINES

A total number of 142 clinical trials are currently being conducted or have been completed in a wide spectrum of tumors, in order to test the effect of several nanoparticles in combination with chemotherapeutic and biological drugs (https://clinicaltrials.gov/). We have used the terms "nanoparticles and cancer therapy" when looking for this information.

From all these studies, 108 use a paclitaxel albumin-stabilized nanoparticle formulation, called Nab-paclitaxel. Nab-paclitaxel is a novel Cremophor® EL-free, non-crystalline, amorphous, albumin-bound nanoparticle formulation of paclitaxel suspended in normal saline. Nab paclitaxel was approved by the US FDA in 2005 to treat breast cancer. Paclitaxel is a taxane that acts as antimicrotubule agent and possess excellent antitumor effects in several types of cancer [233]. However, a main inconvenient for the human use is the high hydrophobicity. Hence, it has been de-scribed that albumin-bound paclitaxel nanoparticles induce lower toxicities in comparison with other taxanes improving both efficacy and safety. As result, this nanoparticle formulation can be used in combination with other chemotherapeutics or therapeutics regimes. In fact, when chemotherapy or radiation is combined with such nanoparticles, the composition is more effective than other drugs combinations [234].

Most clinical trials using Nab-paclitaxel are applied to breast cancer as, for instance, the recently completed clinical trial identified as NCT00675259. This is a phase II trial in-vestigating the side effects and how well administration of Nab-paclitaxel and carboplatin, together with the monoclonal antibody bevacizumab, works in treating women undergoing surgery for stage II or III breast cancer. The results show that neoadjuvant chemotherapy (NCT) with weekly Nab-paclitaxel and carboplatin, and twice per week bevacizumab, resulted in a pathologic complete response (pCR) rate and, consequently, long-term outcomes. These results were superior to the historical data with anthracycline/taxane-containing NCT and to carboplatin and taxane combinations in breast cancer patients HER2– [235].

Another active phase II trial is ongoing to determine the activity of Nab-paclitaxel plus lapatinib, an oral, reversible inhibitor of epidermal growth factor receptor (EGFR) and HER2 tyrosine kinase, in the first- and second-line setting in women with HER2+ metastatic breast cancer (MBC) (NCT00709761). The initial results show a clinical benefit of this combination with respect to investigator-assessed overall response rate in 32 patients (53%).

The highest number of responses was partial response (28 patients; 47%). Four patients (7%) had a complete response and 10 patients (17%) showed disease stability. Moreover, the incidence and severity of adverse events for the combination treatment were in concordance with the safety profiles of lapatinib and Nab- paclitaxel. These preliminary results suggest high efficacy and no further adverse effects [236].

Some examples of other indications for Nab-paclitaxel include: i) melanoma that cannot be removed by surgery used in combination with Bevacizumab and/or Ipilimumab (an anti-cytotoxic T-lymphocyte-associated antigen-4 mono- clonal antibody) (NCT02158520, NCT00626405) or with carboplatin (NCT00404235); ii) patients with advanced pancreatic cancer do not responders to first-line therapy with gemcitabine (NCT00691054), in combination with Gemcitabine and/or radiation (NCT02427841), with sorafenib and virinostat (NCT02349867) or with certinib and cisplatin (NCT02227940) among others; iii) patients with non-small- cell lung cancer (NSLCC) previously treated with chemo- therapy (NCT01620190), in combination with chemotherapy (NCT00073723, NCT01872403, NCT02503358, NCT02016209, NCT00729612) and/or radiotherapy (NCT00544648, NCT00553462); iv) head and neck cancer followed by or together with chemoradiation (NCT01847326, NCT00851 877, NCT00736619).

In addition, there are clinical trials based on other nanoparticle formulations. For example, BIND-014 are do-cetaxel nanoparticles composed of PLA-PEG and functional- ized with a targeting ligand for injectable suspension (WO2013044219 A1) that are being evaluated in patients with advanced solid tumors, including prostate cancer, head and neck carcinoma or NSLCC (NCT01103791, NCT01812746, NCT02479178, NCT01300533, and NCT01792479). Other trials include NBTXR3 crystalline nanoparticles as intratumor injection, which are activated by radiation to generate electrons to destroy the cancer cells causing less damage to the healthy tissues. These 50-nm- sized sphere nanoparticles, such as hafnium oxide are functionalized with a negative surface, have been developed as selective radio enhancers [237], and represent a break- through approach for the local treatment of solid tumors such as soft tissue sarcomas and head and neck cancer (NCT01946867, NCT02379845, NCT01433068). Other nanopharmaceutical formulation, called CRLX101, has been designed consisting of cyclodextrin-based polymer molecule and camptothecin, and appears to concentrate and slowly release camptothecin in tumors over an extended period. CRLX101 inhibits tumor proliferation and angiogenesis showing high anti-tumor potential in heavily pre-treated or treatment-refractory solid-tumor [238]. Studies include the combination with standard neoadjuvant therapies capecitabine and radiation therapy to determine the maximum dose tolerated (phase Ib) and both disease-free survival and over- all survival (phase II) in patients with rectal cancer (NCT02010567), the combination with bevacizumab for recurrent ovarian/tubal/peritoneal cancer (NCT01652079) or the use alone in locally advanced or metastatic NSCLC (NCT01380769).

An ongoing clinical development includes a nanopharmaceutical formulation based on antibody-linked nanoparticles by attaching monoclonal antibodies or antibody frag- ments to the surface of liposomes (immunoliposomes), this enabling more efficient selective binding to antigens expressed in the target cells and their internalization. This study employs anti-EGFR-immunoliposomes to selectively deliver doxorubicin to EGFR-overexpressing tumor cells in solid tumors (NCT01702129).

Other active clinical trials use micellar nanoparticles such as padical (NCT00989131), NK105 (NCT01644890) or NC-6004 (NCT02240238), minicells containing a microRNA mimic (NCT023691989), cyclodextrin PEG-based nanoparticle containing siRNA (CALAA-01) (NCT00689065) or carbon nanoparticles (NCT02123407).

Several nano-formulated drugs have been evaluated, of which a few have already been approved for clinical use (Table 2) while others are undergoing phase 2/3 clinical trials. This is due to their potential for toxicity to patients and to the environment, the high cost of production, and premature drug release. Conversely, they have the ability to convert insoluble or poorly soluble drugs, avoiding toxic organic solvents in addition to decreasing the therapeutic dosage of conventional chemotherapy.

CONCLUSION

In this review, we have evidenced the tremendous potential that smart drug-delivery systems such as nanoparticles have to enhance the therapeutic effect of current standard treatment modalities, including chemotherapies and radio- therapies. Increased numbers of nanoparticles are been developed and studied for their anti-cancer properties. Recent ongoing clinical trials add modifications for the selective targeting of cancer cells that will further improve the quality of life of cancer patients, by reducing the adverse effects of chemotherapeutic agents, and enhancing overall survival. Although clinical application of nanotechnology has only just started, additional studies are needed due to the great variability in growth and inner structure of tumors, including the microenvironment. Therefore, a better understanding of the barriers that keep efficacy low, preventing uniform delivery of nanoparticles into tumors is needed in order to develop strategies for further improvement [239]. In addition, the adverse secondary effects range from inert to highly toxic due to the ample variety of nanomaterials available.

Moreover, the increasingly widespread availability of affordable technologies has allowed a deeper understanding of cancer at the genetic level, leading to new paradigms that are improving patient outcomes and helping in treatment decisions [8, 233, 240]. Since the molecular profiling of tumors in each person is unique, the selection of a specific targeted agent or a specific treatment regimen based on its individual molecular profile is generating great expectations both to oncologists as well as to patients. These characteristics make nanoparticles highly appropriate therapeutic strategy with application in targeted, personalized medicine, where they could transport large amount of drugs specifically into malignant cells avoiding undesired effects in healthy cells. This new era of individualized or stratified medicine allows that the right therapeutic regime is given to each patient speeding up clinical trials and, finally, resulting in the improvement of cancer patients [241].

In addition, the development of novel nanoplatform-targeted therapies against cancer stem cells, which are involved in chemoresistance, recurrence of the disease, and even on metastasis formation [242], will improve the clinical outcome for cancer patients.

To have conclusive results concerning ongoing phase-I clinical trials assessing the safety and pharmacokinetics of nanoparticles formulations is vital in order to standardize the application of treatments based on smart drug-delivery systems to the daily clinical practice.

CONFLICT OF INTEREST

The authors confirm that this article content has no conflict of interest.

ACKNOWLEDGEMENTS

This work was supported by the projects MAT2013-43922-R, MAT2015-63644-C2-2-R, (European FEDER support included, MINECO Spain), PI-0533-2014 (Fun- dación Progreso y Salud/Janssen Cilag, Junta de Andalucía, Spain) and P07-FQM2496, P10-CTS-6270 and P07- FQM3099 (Junta de Andalucía, Spain).

TABLES AND FIGURES

Table 1. Examples of different nanostructured carrier systems and the properties that influence their suitability for parenteral administration.

| Nanocarrier | Size Range | Solubility | Stability | Clearance/Filtration | Toxicity | Uptake by Cancer Cells | Suitability for Parenteral Administration | Ref. |
|---|---|---|---|---|---|---|---|---|
| Liposomes | 50nm-above 1µm | High solubility | Poor stability | Rapid removal by reticuloendothelial system | Almost biologically inert. Can cause "complement activation-related pseudoallergy" | High uptake | *** | [81] [202] |
| Polimeric Nanoparticles | 1-1000nm | High solubility | Stable | Long time circulation in blood | Reduced toxicity with a limited interaction with healthy cells | Enhanced uptake | *** | [81] |
| Lipid Nanoemulsions | less than ~200–300 nm | High solubility | kinetically stable | The use of polymers with hydrophilic segment such as PEG, poloxamers, or chitosan helps to extend the life-time in the bloodstream | Low toxicity and reduced side effects | Nanoparticles smaller than 100 nm preferentially accumulate in cancer tissues. High uptake | *** | [81] |
| Solid Lipid Nanoparticles | 40-1000nm | Poor stability | High physical stability | Optimum particle size for lymphatic uptake 10-100 nm. Increase in particle size leads to extended blood circulation | Good tolerability and biodegradability | Relatively low tumor uptake | * | [15] |
| Ceramic nanoparticles | 2-1000nm | Generally soluble but TiO2 NPs | High stability | Efficient renal clearance < 10nm. Liver, spleen accumulation and hepatic excretion if >10nm | Generally bio-inert but non biodegradable. Production of ROS for TiO2. | Higher below 200nm | ** | [143] [203] [204] |
| Carbon nanotubes | 1-50 nm diameter 100nm-5µm in length | Poorly soluble. Solubility improved with chemical modifications | Native aggregated state. Processing strategies to modify the surface commonly practiced | A higher degree of surface functional groups causes lesser accumulation in the liver and faster clearance through the renal system. Rapid clearance of <500 nm NPs | Highly cytotoxic; asbestos-like toxicity. Biocompatibility increases if functionalized with soluble and biocompatible materials | Acceptable tumor uptake | * | [81] [205] |
| Gold nanoparticles | 2-400nm | High. Few cases of hemolysis and blood clotting | High | 5-30nm clearance by kidney. >40nm accumulation in liver and spleen | Very low. But production of ROS | Very high for NPs below 100nm | *** | [206-208] |
| Iron oxide nanoparticles | 5-1000nm | Coating makes them highly hydrophilic | Excellent chemical stability | Poor. Accumulation in liver and spleen | Low | High for small particles | *** | [209, 210] |

*** very suitable; ** suitable; * low suitable.

Table 2. Nanomedicines currently available on the market.

| Product | Company | Drug | Formulation | Indication | Route of Administration |
|---|---|---|---|---|---|
| Abraxane | Abrasix Bioscience, AstraZeneca<br>Celgene | Paclitaxel<br>Paclitaxel in combination with gemcitabine | Albumin-bound nanoparticles | Metastatic breast cancer<br>Metastatic pancreatic cancer | iv[a] |
| DaunoXome | Gilead Science | Daunorubicin | Liposome | HIV-related Kaposi sarcoma | iv[a] |
| DepoCyt | Skype Pharma Enzon | Cytarabine | Liposome | Lymhomatous meningitis | intrathecal |
| Doxil | Sequus Pharmaceutical | Doxorubicin | Liposome | Kaposi sarcoma | iv[a] |
| Doxy (Caelyx) | Orthobiotech Schering-Plough Janssen | Doxorubicin | Pegylated liposome | Metastatic breast and ovarian cancer; Kaposi sarcoma | im[a] |
| Genexol-PM | Samyang | Paclitaxel | Polymeric micelle | Breast cancer, Small cell lung cancer | iv[a] |
| Marqibo | Talon therapeutics | Vincristine | Liposome | Philadelphia chromosome-negative lymphoblastic leukemia | iv[a] |
| Mepact (mifamurtide) | IDM Pharma SAS | Synthetic derivative of muramyl dipeptide | Liposome | Osteosarcoma | iv[a] |
| Myocet | Elan Pharmaceuticals / Sopherion Therapeutics Zeneus Pharma Ltd | Doxorubicin | Liposome | Metastatic breast cancer | iv[a] |
| Oncaspar | Enzon | Asparaginase | PEG nanoparticle | Acute Lymphocytic Leukemia | iv/im[a] |
| Zinostatin Stimalmer | Yamanouchi Japan | Neocarzinostatin | Polymer-drug conjugates | Hepatocellular carcinoma | Local, via hepatic artery infusion |

[a] iv, intravenous; im, intramuscular.

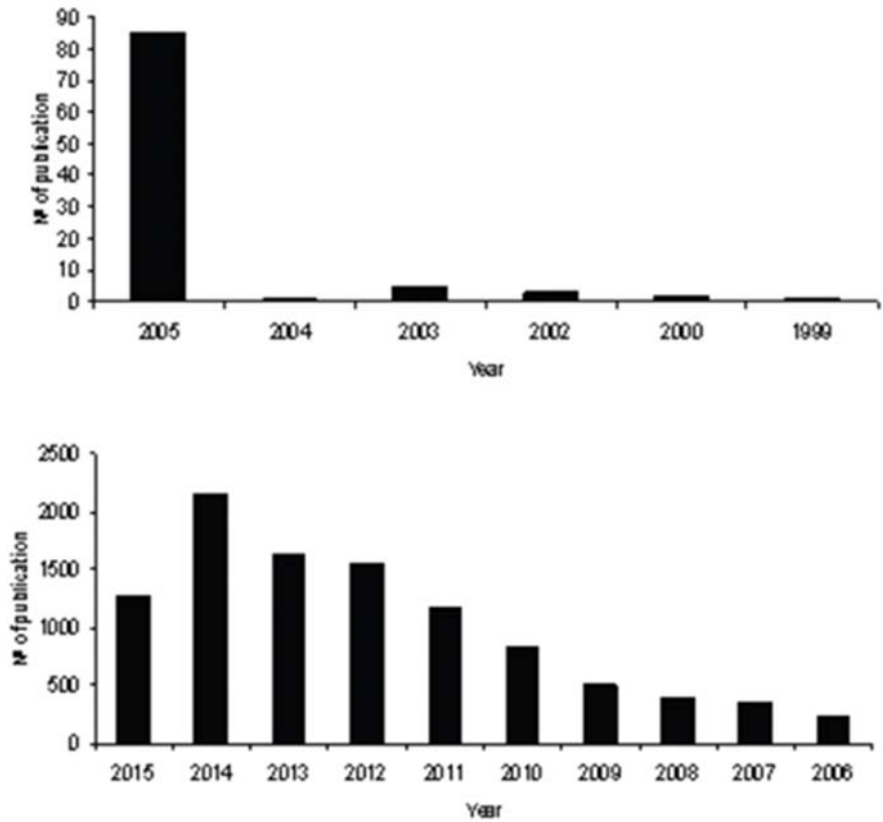

Fig. (1). Time course of publications related to nanomedicine. Data reported here reflect manuscripts available through PubMed database until September 2015 using the term "nanomedicine" as a key word. (http://www.ncbi.nlm.nih.gov/pubmed/?term=nanomedicine).

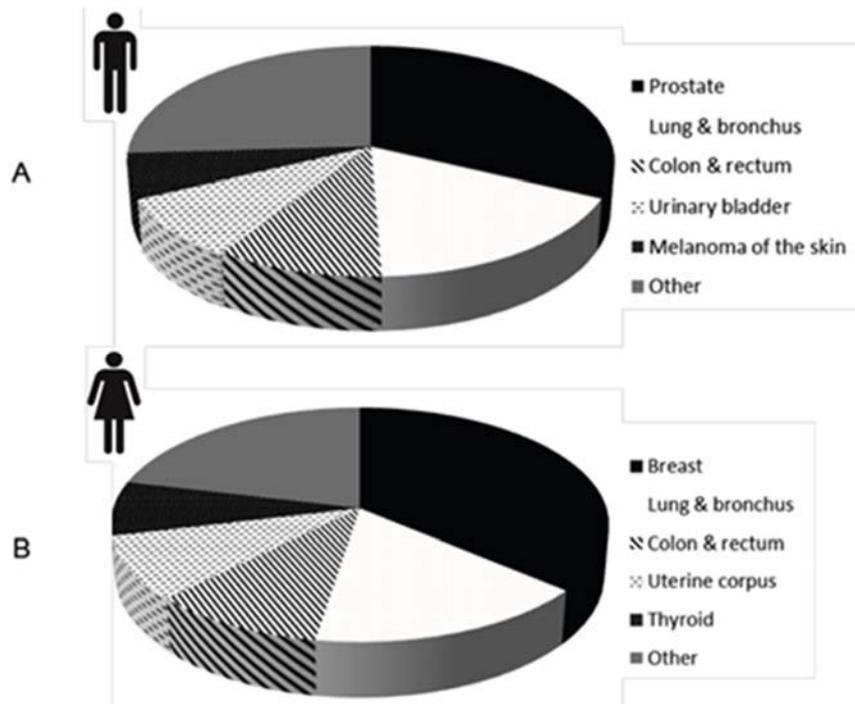

Fig. (2). Five leading cancer types for the estimated deaths, by gender in the United States.

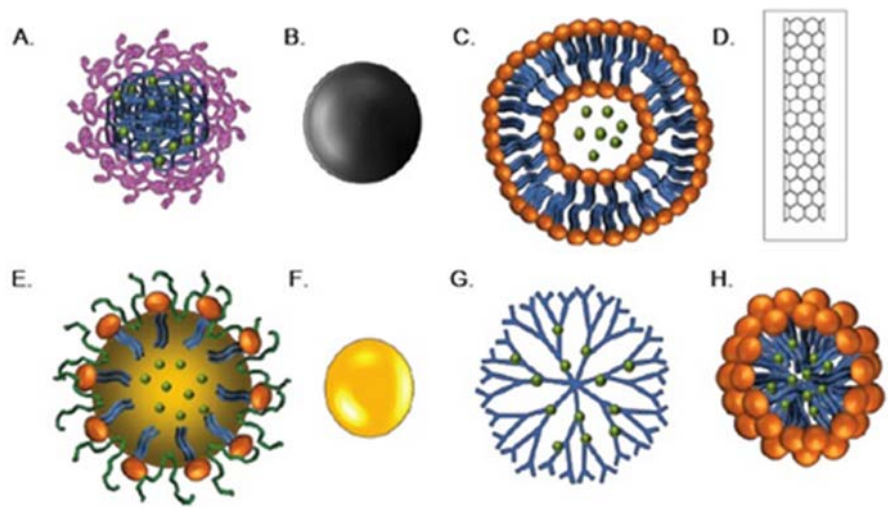

Fig. (3). Examples of drug-delivery platforms: A) Polymeric nanoparticle. B) Magnetic nanoparticle. C) Liposome. D) Carbon nanotube. E) Nanoemulsion. F) Gold nanoparticle. G) Dendrimer. H) Micelle.

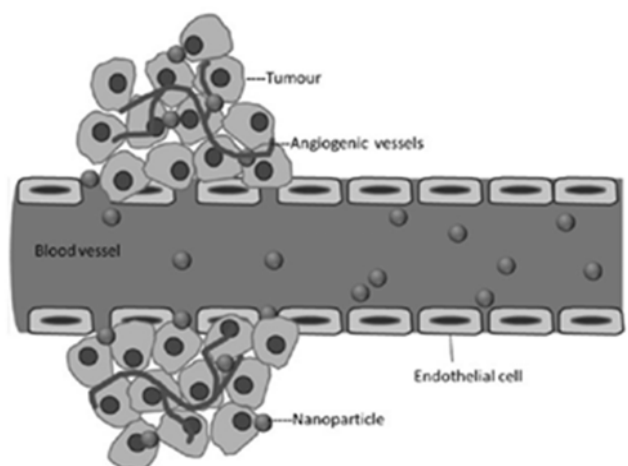

Fig. (4). Schematic representation of passive tissue targeting, achieved by tumor-site accumulation of nanoparticles that extravasate through enlarged pores of tumor vasculature (EPR effect).

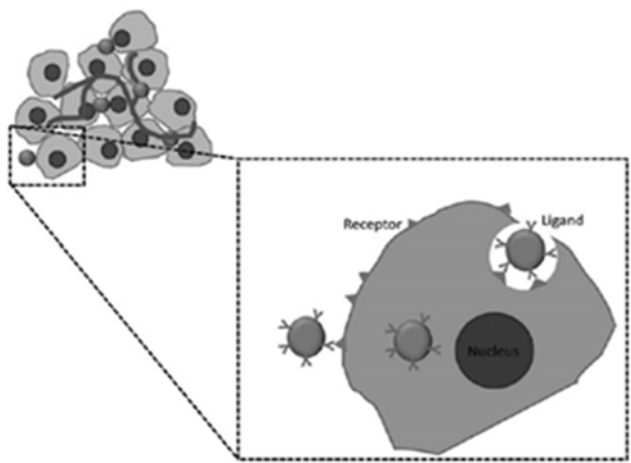

Fig. (5). Schematic representation of active targeting which can be achieved by functionalizing the surface of nanocarriers with moie- ties that specifically recognize and bind to tumor cells.

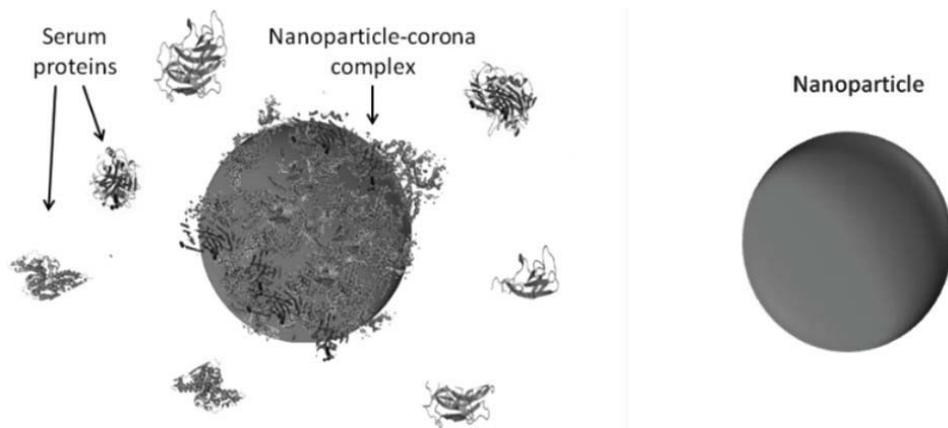

Fig. (6). Schematic representation of protein corona developed on the surface of nanoparticles in the presence of serum proteins.

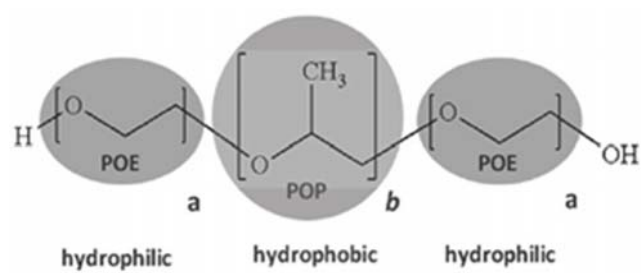

Fig. (7). Structure of Pluronics.